\documentclass{article}
\usepackage{amssymb}
\usepackage{amsmath}

\newtheorem{theorem}{Theorem}

\newtheorem{proposition}[theorem]{Proposition}

\input{tcilatex}
\begin{document}

\title{\textbf{Constructing the CKVs of Bianchi III and V spacetimes}}
\author{Antonios Mitsopoulos$^{1}$\thanks{%
Email:antmits@phys.uoa.gr}, Michael Tsamparlis$^{1}$\thanks{%
Email: mtsampa@phys.uoa.gr} and \ Andronikos Paliathanasis$^{2}$\thanks{%
Email: anpaliat@phys.uoa.gr} \\
{\ \ }\\
$^{1}${\textit{Faculty of Physics, Department of
Astronomy-Astrophysics-Mechanics,}}\\
{\ \textit{University of Athens, Panepistemiopolis, Athens 157 83, Greece.}}%
\\
\\
$^{2}${\textit{Institute of Systems Science, Durban University of Technology}%
}\\
{\textit{Durban 4000, Republic of South Africa.}}}
\date{}
\maketitle

\begin{abstract}
We determine the conformal algebra of Bianchi III and Bianchi V spacetimes
or, equivalently, we determine all Bianchi III and Bianchi V spacetimes
which admit a proper conformal Killing vector. The algorithm that we use has
been developed in Class. Quantum. Grav. 15, 2909 (1998) and concerns the
computation of the CKVs of decomposable spacetimes. The main point of this
method is that a decomposable space admits a CKV if the reduced space admits
a gradient homothetic vector the latter being possible only if the reduced
space is flat or a space of constant curvature. We apply this method in a
stepwise manner starting from the two dimensional spacetime which admits an
infinite number of CKVs and we construct step by step the Bianchi III and V
spacetimes by assuming that CKVs survive as we increase the dimension of the
space. We find that there is only one Bianchi III and one Bianchi V
spacetime which admit at maximum one proper CKV. In each case we determine
the conformal Killing vector and the corresponding conformal factor. As a
first application in these two spacetimes we study the kinematics of the
comoving observers and the dynamics of the corresponding cosmological fluid.
As a second application we determine in these spacetimes generators of the
Lie symmetries of the wave equation.

\end{abstract}

\section{Introduction}

\label{intro}

A conformal Killing vector (CKV) $\mathbf{X}$ of a metric $g_{ab}$ is a
vector field that satisfies the condition $L_{\mathbf{X}}g_{ab}=2\psi g_{ab}$
where $\psi (x^{r})$ is the conformal factor. The CKVs are classified as
Killing Vectors (KVs) for $\psi =0$; Homothetic Vectors (HVs) for $\psi
=const$; Special Conformal Killing Vectors (SCKVs) for $\psi _{;ab}=0$; and
proper CKVs for $\psi _{;ab}\neq 0$. The HVs and the KVs are also non-proper
Affine Collineations and non-proper Projective Collineations.

The knowledge of the proper CKVs of a given spacetime is important because
they act as geometric constraints which can be used in the study of the
kinematics and the dynamics of a given spacetime. For example a CKV can be
used to reduce the number of unknowns of a gravitational (or cosmological)
model and also to increase the possibility of finding new solutions of
Einstein's field equations (see for example \cite{Maart}, \cite{Coley1992}, 
\cite{Coley1994}, \cite{Maa-Mah-Tup}, \cite{EMSF}, \cite{ct01}, \cite{ct02}, 
\cite{ct03}, \cite{ct04}). Furthermore the conformal algebra can be used in
order to classify spaces (e.g. Finsler manifolds, pseudo-Euclidean
manifolds) (see \cite{Frances}, \cite{Johar}). For example one may use the
CKVs of a space in order to determine the classes of manifolds which are
conformally related to the given space; or to use them in order to study the
locally conformal flatness of a space around a singularity (i.e. a point $%
x_{0}$ where the CKV vanishes). The special class of the CKVs, the KVs, have
been used in numerous applications. For example Geroch in \cite{Geroch1} and 
\cite{Geroch2} has used the KVs in order to derive new solutions of the
gravitational field equations. The gradient KVs can be used also to
decompose the spacetime metric and simplify the field equations. Moreover,
KVs are related with the conservation laws for the geodesic equations,
indeed KVs form a subalgebra on the Noether symmetries for the geodesic
Lagrangian, for more details see \cite{mtsgeo}

Apart of the above applications, another important area where the CKVs and
the more general projective collineations have been used is the geometric
study of Lie symmetries of differential equations. For example early studies
of the geodesic equations \cite{katz1}, \cite{katz2}, \cite{ami1}, \cite%
{ami2} have shown a unique connection of the Lie point symmetries for the
geodesic equations in a space with the elements of the projective algebra of
the space. Furthermore in a conservative dynamical system one may consider
the kinetic energy as a metric and study the Lie and the Noether point
symmetries of the dynamic equations using the collineations of this metric.
In \cite{p1}, \cite{p2} it has been shown that in these case the Lie point
symmetries are generated by the special projective algebra and the Noether
point symmetries by the Homothetic algebra of the kinetic metric. Similar
results have been found for some partial differential equations of special
interest in curved spacetimes, as the wave and the heat equation (see \cite%
{p5a}, \cite{p4}, \cite{p5} and references therein) where it has been shown
that in these cases the Lie point symmetries involve the CKVs.

In the present work we apply the propositions and the methodology developed
in \cite{Tsamp-Nikol-Apost}, \cite{Tsamp-Apost} and \cite{bickv} in order to
determine all Bianchi III and V spacetimes that admit proper Conformal
Killing Vectors (CKVs). The Bianchi I\ spacetimes which admit proper CKVs
have been determined in \cite{bickv}. Bianchi III and Bianchi V spacetimes
are of special interest and they find many applications in the study of
anisotropic cosmologies. Bianchi can provide a different cosmological
behaviour in the early universe \cite{tch1}, while also they can been seen
as the homogeneous limits of exact inhomogeneous cosmological models \cite%
{kra}.

Bianchi spacetimes are spatially homogeneous spacetimes of the general form 
\begin{equation}
ds^{2}=-dt^{2}+A^{2}(t)(\omega _{1})^{2}+B^{2}(t)(\omega
_{2})^{2}+C^{2}(t)(\omega _{3})^{2}  \label{ds.bianchi}
\end{equation}%
where $\omega _{i}$, $i=1,2,3$, are basis 1-forms and $A(t)$, $B(t)$, $C(t)$
are functions of the time coordinate (see \cite{Stephani}, \cite{Wald}, \cite%
{Ryan}). For instance, 
\begin{eqnarray*}
\text{Bianchi I} &:&\omega _{1}=dx~,~\omega _{2}=dy~,~\omega _{3}=dz \\
\text{Bianchi III} &\text{:}&\omega _{1}=dx~,~\omega _{2}=dy~,~\omega
_{3}=e^{-x}dz \\
\text{Bianchi V} &\text{:}&\omega _{1}=dx~,~\omega _{2}=e^{x}dy~,~\omega
_{3}=e^{x}dz.
\end{eqnarray*}%
In the case of $B^{2}(t)=C^{2}(t)$ the Bianchi spacetimes contain a fourth
isometry which is the rotation of the $yz$ plane and reduce to the important
subclass of Locally Rotational Symmetric (LRS) spacetimes (see for example 
\cite{Tsamparlis2001} and citations therein).

\textbf{\ }The structure of the paper is as follows. In section \ref{sec2}
we briefly discuss the method we apply for the determination of the proper
CKVs of a decomposable spacetime. Sections \ref{sec3} and \ref{sec4} contain
the main calculations of the paper and the propositions derived from the
application of the method mentioned in section \ref{sec2}. In section \ref%
{sec5} we apply our results in the cases of Bianchi III and Bianchi V
cosmological models. Section \ref{sec6} contains an application of the CKVs
on the symmetries of differential equations. More specifically, we compute
the generic Lie point symmetries of the wave equation for the Bianchi
spacetimes constructed in sections \ref{sec3} and \ref{sec4}. Finally, in
section \ref{sec7} we draw our conclusions.

\section{Preliminaries}

\label{sec2}

We recall that a Riemannian manifold is decomposable along the coordinate $t$
iff the metric $g_{ab}$ admits the non-null gradient KV $u^{a}=\partial t$
where $u^{a}u_{a}=\varepsilon u^{2}$ , $\varepsilon =\pm 1.$ In this case
one defines the projection operator 
\begin{equation}
h_{ab}=g_{ab}-\frac{\varepsilon }{u^{2}}u_{a}u_{b}  \label{f1}
\end{equation}%
and decomposes the tensor algebra along $u^{a}$ and normal to $u^{a}.$ For $%
n-$dimensional decomposable Riemannian manifolds $M^{n}$ with $n\geq 3$ an
algorithm has been developed (see \cite{Tsamp-Nikol-Apost}) which determines
the proper CKVs in terms of the (gradient) proper CKVs of the $%
\left(n-1\right) $ non-decomposable space.

In particular, it has been shown that an $n$-dimensional decomposable space $%
M^{n}$ admits proper CKVs if and only if the $\left( n-1\right) $ space $%
M^{n-1}$ admits a gradient proper CKV whose conformal factor is the gradient
factor which constructs the (gradient) CKV. In addition, any gradient proper
CKV of the $M^{n-1}$ provides two proper CKVs for the $M^{n}$. For a four
dimensional manifold the following result is shown in \cite%
{Tsamp-Nikol-Apost}.

If $M^{n}$, $n=4$ is a decomposable Riemannian manifold with line element ($%
\mu ,\nu =1,2,3)$%
\begin{equation}
ds^{2}=\varepsilon dt^{2}+h_{\mu \nu }\left( x^{\sigma }\right) dx^{\mu
}dx^{\nu }  \label{sx2.1}
\end{equation}%
where $\varepsilon =\pm 1,$ the vector field 
\begin{equation}
X^{a}\partial _{a}=-\frac{\varepsilon }{p}\dot{\lambda}\left( t\right) \psi
\left( x^{\sigma }\right) \partial _{t}+\frac{1}{p}\lambda \left( t\right)
\xi ^{\mu }\left( x^{\sigma }\right) \partial _{\mu }+L^{\mu }\partial _{\mu
}  \label{sx2.0a}
\end{equation}%
is a proper CKV of (\ref{sx2.1}) where

a. $L^{\mu }$ is a non-gradient KV or HV of $M^{n-1}$

b. $\xi ^{\mu }\left( x^{\sigma }\right) $ is a gradient proper CKV of $%
M^{n-1}$ with conformal factor $\psi \left( x^{\sigma }\right) $ i.e. $%
\mathit{L}_{\xi }h_{\mu \nu }\left( x^{\sigma }\right) =2\psi \left(
x^{\sigma }\right) h_{\mu \nu }\left( x^{\sigma }\right) $

c. The function $\lambda \left( t\right) $ is given by 
\begin{equation}
\lambda \left( t\right) =\lambda _{1}e^{i\sqrt{\varepsilon p}t}+\lambda
_{2}e^{-i\sqrt{\varepsilon p}t},\enskip\text{for $\varepsilon p>0$}
\label{sx2.2}
\end{equation}%
or 
\begin{equation}
\lambda \left( t\right) =\lambda _{1}e^{\sqrt{-\varepsilon p}t}+\lambda
_{2}e^{-\sqrt{-\varepsilon p}t},\enskip\text{for $\varepsilon p<0$}
\label{sx2.2b}
\end{equation}%
where $p$ is a non-vanishing constant and $\lambda _{1}$, $\lambda _{2}$ are
independent constants, provided the function $\psi \left( x^{\sigma }\right) 
$ satisfies the condition 
\begin{equation}
\psi _{;\mu \nu }=p\psi h_{\mu \nu }.  \label{sx2.1a}
\end{equation}

Concerning the homothetic vector it has been shown in \cite%
{Tsamp-Nikol-Apost} that when the $M^{n-1}$ space admits a HV $H^{\mu
}\left( x^{\sigma }\right) $ with conformal factor $C$, the $M^{n}$ admits
the HV 
\begin{equation}
H^{a}\partial _{a}=Ct\partial _{t}+H^{\mu }\partial _{\mu }.  \label{sx2.3}
\end{equation}%
Finally, concerning the Killing vector fields, it has been shown that the
Killing vector fields of $M^{n}$ are 
\begin{equation}
K^{a}=k_{0}\partial _{t}+k_{1I}h^{\mu \nu }\left( x^{\sigma }\right) K_{\nu
}^{I}\left( x^{\sigma }\right) +k_{2I}h^{\mu \nu }\left( x^{\sigma }\right)
S_{,\nu }^{I}\left( x^{\sigma }\right) +k_{3I}\left( -\varepsilon
S^{I}\left( x^{\sigma }\right) \partial _{t}+h^{\mu \nu }\left( x^{\sigma
}\right) S_{,\nu }^{I}\left( x^{\sigma }\right) \right)  \label{sx2.4}
\end{equation}%
where $K_{\nu }^{I}\left( x^{\sigma }\right) $ are the non-gradient KVs of $%
M^{n-1}$ and $S_{,\nu }^{I}\left( x^{\sigma }\right) $ are the gradient KVs
of $M^{n-1}$. Finally, $k_{0}$, $k_{1I}$, $k_{2I}$ and $k_{3I}$ are
independent constants.

However another possibility that the space (\ref{sx2.1}) admits proper CKVs
is when it is conformally flat. That case was found to be important in the
classification of Bianchi I spacetimes in \cite{bickv} according to the
admitted CKVs, but it does not provide any result in the case of Bianchi III
and Bianchi V spacetimes, thus we omit it from the present discussion.

The concept of conformally related metrics plays a crucial role in the
computation of the CKVs, therefore we review the basic definitions
concerning these metrics. Two metrics $\widehat{g}_{ab},$ $g_{ab}$ are said
to be conformally related iff there is a function $N^{2}(x^{r})$ such that $%
\widehat{g}_{ab}=N^{2}(x^{r})g_{ab}$. The conformally related metrics share
the same conformal algebra but with different conformal factors. For a given
vector field $\mathbf{X}$ we have the decompositions/identities 
\begin{equation*}
L_{\mathbf{X}}\widehat{g}_{ab}=2\widehat{\psi }(\mathbf{X})\widehat{g}_{ab}+2%
\widehat{H}_{ab}(\mathbf{X})\enskip\text{and}\enskip L_{\mathbf{X}%
}g_{ab}=2\psi (\mathbf{X})g_{ab}+2H_{ab}(\mathbf{X}).
\end{equation*}%
where $H_{ab}(\mathbf{X}),\widehat{H}_{ab}(\mathbf{X})$ are symmetric
traceless tensors. It can be shown that 
\begin{equation*}
\widehat{\psi }(\mathbf{X})=\mathbf{X}(\ln N)+\psi (\mathbf{X}),\enskip%
\widehat{H}_{ab}(\mathbf{X})=N^{2}H_{ab}(\mathbf{X})
\end{equation*}%
and 
\begin{equation*}
\widehat{F}_{ab}(\mathbf{X})=N^{2}F_{ab}(\mathbf{X})-2NN_{,[a}X_{b]}
\end{equation*}%
where $\widehat{F}_{ab}(\mathbf{X})=\widehat{X}_{[a\hat{;}b]}=\widehat{X}%
_{[a;b]}$ and $F_{ab}(\mathbf{X})=X_{[a;b]}$. Moreover 
\begin{eqnarray*}
\widehat{X}_{a\hat{;}b} &=&\frac{1}{2}L_{\mathbf{X}}\widehat{g}_{ab}+%
\widehat{F}_{ab}(\mathbf{X}) \\
X_{a;b} &=&\frac{1}{2}L_{\mathbf{X}}g_{ab}+F_{ab}(\mathbf{X}).
\end{eqnarray*}

A metric $g_{ab}$ is called conformally flat iff it is conformal to the flat
metric $\eta _{ab}$. A metric conformally related to a conformally flat
metric is also conformally flat. It is well known that all the 2d-spacetimes
are conformally flat and admit an infinity number of CKVs while only the
flat 2d-metrics admit special CKVs.

The flat $n$-dimensional metric $\eta _{ab}$ admits an algebra of $\frac{%
(n+1)(n+2)}{2}$ CKVs which consists of $n+\frac{n(n-1)}{2}=\frac{n(n+1)}{2}$
KVs, one HV and $n$ proper SCKVs. The generic (not proper!) SCKV of a flat
metric is given by the formula%
\begin{equation}
\chi ^{a}\partial _{a}=\alpha ^{a}\mathbf{P}_{a}+\alpha ^{BA}\mathbf{r}%
_{AB}+\beta \mathbf{H}+2\beta ^{a}\mathbf{K}_{a}  \label{eq.flatCKV1}
\end{equation}%
with conformal factor $\psi =\beta +2\beta _{a}x^{a}$ and integration
constants $\alpha ,\beta ,\alpha ^{a},\beta ^{a}$, $\alpha _{ab}=-\alpha
_{ba}$. In this expression $\mathbf{P}_{a}$ are the $n$ KVs (translations), $%
\mathbf{r}_{AB}$ are the $\frac{n(n-1)}{2}$ KVs (rotations), $\mathbf{H}$ is
the HV (dilatation) and $\mathbf{K}_{a}$ are the $n$ proper SCKVs. The
summation over $A$, $B$ satisfies the condition $1\leq A<B\leq n$.

In a coordinate system in which the metric has its reduced form $\eta
_{ab}=diag(-1,...,-1,+1,...,+1)$ the above vectors are given by the
following expressions $\mathbf{P}_{a}=\delta _{a}^{b}\partial _{b},\mathbf{r}%
_{ab}=2\delta _{\lbrack a}^{c}\delta _{b]}^{d}x_{c}\partial _{d},$ $\mathbf{H%
}=x^{a}\partial _{a}$ and $\mathbf{K}_{a}=\left( x_{a}x^{b}-\frac{1}{2}%
\delta _{a}^{b}x_{c}x^{c}\right) \partial _{b}=x_{a}\mathbf{H}-\frac{1}{2}%
(x_{b}x^{b})\mathbf{P}_{a}.$ These vector fields span the conformal algebra
of the flat space $\eta _{ij}$.

\section{CKVs of Bianchi III spacetime}

\label{sec3}

Consider the three-dimensional decomposable spacetime of Lorentzian signature%
\begin{equation}
ds_{(1+2)}^{2}=\Gamma ^{2}\left( \tau \right) \left( -d\tau
^{2}+dx^{2}\right) +dy^{2}.  \label{b3.01}
\end{equation}%
The line element (\ref{b3.01}) for arbitrary $\Gamma \left( \tau \right) $
admits a two-dimensional conformal Killing algebra consisting by the KVs $%
\partial _{y}$ and $\partial _{x}$.

For the conformal spacetime 
\begin{equation}
d\bar{s}_{\left( 1+2\right) }^{2}=B^2\left( \tau \right) e^{2x}ds_{(1+2)}^{2}
\label{b3.02}
\end{equation}%
the vector field $\partial _{y}$ remains a KV but $\partial _{x}$ now
becomes a proper HV.

Consider now the four-dimensional decomposable spacetime 
\begin{equation}
ds_{\left( 1+3\right) }^{2}=d\bar{s}_{\left( 1+2\right) }^{2}+dz^{2}
\label{b3.03}
\end{equation}%
which admits a three-dimensional conformal algebra consisting of the KVs $%
\partial _{y},~\partial _{z}$ and the proper HV $\partial _{x}+z\partial
_{z} $. Then, the conformally related spacetime $ds_{\left( III\right)
}^{2}=A^{2}\left( \tau \right) e^{-2x}ds_{\left( 1+3\right) }^{2}$ which can
be written equivalently\footnote{%
Where $\alpha ^{2}\left( t\right) =A^{2}\left( t\right) B^{2}\left( t\right)
\Gamma ^{2}\left( t\right) $,~$\beta ^{2}\left( t\right) =A^{2}\left(
t\right) B^{2}\left( t\right) $ and $\gamma ^{2}\left( t\right) =A^{2}\left(
t\right) $ while $t=\int a\left( \tau \right) d\tau $.}%
\begin{equation}
ds_{\left( III\right) }^{2}=-dt^{2}+\alpha ^{2}\left( t\right) dx^{2}+\beta
^{2}\left( t\right) dy^{2}+\gamma ^{2}\left( t\right) e^{-2x}dz^{2}
\label{b3.04}
\end{equation}%
is a Bianchi III spacetime and the vector fields $\partial _{y},~\partial
_{z},~\partial _{x}+z\partial _{z}$ form the Killing algebra of (\ref{b3.04}%
). Therefore, in order the Bianchi III (\ref{b3.04}) to admit greater
conformal algebra the functions $\alpha \left( t\right) ,~\beta \left(
t\right) $ and $\gamma \left( t\right) $ must be specified. Recall that when 
$\alpha \left( t\right) =\gamma \left( t\right) $ spacetime (\ref{b3.04}) is
locally rotational and admits as extra KV the rotation in the two
dimensional space~$ds^{2}=dx^{2}+e^{-2x}dz^{2}$.

The three-dimensional space (\ref{b3.01}) admits a greater conformal algebra
for specific functions $\Gamma \left( \tau \right) $. From the discussion of$%
~$Section \ref{sec2} it follows that $\Gamma \left( \tau \right) $ must be
such that the two-dimensional space 
\begin{equation}
ds_{\left( 2\right) }^{2}=\Gamma ^{2}\left( \tau \right) \left( -d\tau
^{2}+dx^{2}\right)  \label{b3.05}
\end{equation}%
admits proper gradient CKVs or a greater Killing algebra. For
two-dimensional spaces it is well-known that the admitted KVs can be zero,
one or three and in the latter case the space is maximally symmetric. Since (%
\ref{b3.05}) admits always the KV $\partial _{x}$, the $\Gamma \left( \tau
\right) $ must be specified so that (\ref{b3.05}) is maximally symmetric.
Without loss of generality we can select $\Gamma ^{2}\left( \tau \right)
=e^{m\tau }$ in which case (\ref{b3.05}) is the flat space with Ricci Scalar 
$R_{\left( 2\right) }=0$, or $\Gamma ^{2}\left( \tau \right) =\kappa
^{-2}\cos ^{-2}\left( \tau \right) $ in which case $R_{\left( 2\right)
}=2\kappa ^{2}$.

Furthermore, because all the two-dimensional spaces admit infinitely many
CKVs, the requirement that at least one of the proper CKVs is to be
gradient, specifies the spacetime to be of nonzero constant curvature, which
is a maximally symmetric space and it admits five gradient proper CKVs.

\subsection{Case $\Gamma^2 \left( \protect\tau \right) =e^{m \protect\tau}$}

In the case of $\Gamma ^{2}\left( \tau \right) =e^{m \tau}$ the
three-dimensional space 
\begin{equation}
ds_{(1+2)}^{2}=e^{m\tau}\left( -d\tau ^{2}+dx^{2}\right) +dy^{2}.
\label{b3.06}
\end{equation}%
is flat and admits a ten-dimensional conformal algebra. This algebra
consists of the six KVs%
\begin{equation*}
\mathbf{Y}_{1}=\frac{2}{m}e^{-\frac{m}{2}(\tau -x)}\partial _{\tau }-\frac{2%
}{m}e^{-\frac{m}{2}(\tau -x)}\partial _{x}
\end{equation*}%
\begin{equation*}
\mathbf{Y}_{2}=-\frac{2}{m}e^{-\frac{m}{2}(\tau +x)}\partial _{\tau }-\frac{2%
}{m}e^{-\frac{m}{2}(\tau +x)}\partial _{x}
\end{equation*}%
\begin{equation*}
\mathbf{Y}_{3}=\partial _{x}~~,~~\mathbf{Y}_{4}=\partial _{y}
\end{equation*}%
\begin{equation*}
\mathbf{Y}_{5}=ye^{-\frac{m}{2}(\tau +x)}\partial _{\tau }+ye^{-\frac{m}{2}%
(\tau +x)}\partial _{x}+\frac{2}{m}e^{\frac{m}{2}(\tau -x)}\partial _{y}
\end{equation*}%
\begin{equation*}
\mathbf{Y}_{6}=-ye^{-\frac{m}{2}(\tau -x)}\partial _{\tau }+ye^{-\frac{m}{2}%
(\tau -x)}\partial _{x}-\frac{2}{m}e^{\frac{m}{2}(\tau +x)}\partial _{y}
\end{equation*}%
the HV%
\begin{equation*}
\mathbf{Y}_{7}=\frac{2}{m}\partial _{\tau }+y\partial _{y} \enskip ,\enskip %
\psi_{(1+2)}(\mathbf{Y}_{7})=1
\end{equation*}%
and the three special CKVs 
\begin{equation*}
\mathbf{Y}_{8}=\left[ \frac{2}{m^{2}}e^{\frac{m}{2}(\tau -x)}+\frac{y^{2}}{2}%
e^{-\frac{m}{2}(\tau +x)}\right] \partial _{\tau }+\left[ -\frac{2}{m^{2}}e^{%
\frac{m}{2}(\tau -x)}+\frac{y^{2}}{2}e^{-\frac{m}{2}(\tau +x)}\right]
\partial _{x}+\frac{2y}{m}e^{\frac{m}{2}(\tau -x)}
\end{equation*}%
\begin{equation*}
\mathbf{Y}_{9}=-\left[ \frac{2}{m^{2}}e^{\frac{m}{2}(\tau +x)}+\frac{y^{2}}{2%
}e^{-\frac{m}{2}(\tau -x)}\right] \partial _{\tau }+\left[ -\frac{2}{m^{2}}%
e^{\frac{m}{2}(\tau +x)}+\frac{y^{2}}{2}e^{-\frac{m}{2}(\tau -x)}\right]
\partial _{x}-\frac{2y}{m}e^{\frac{m}{2}(\tau +x)}
\end{equation*}%
\begin{equation*}
\mathbf{Y}_{10}=my\partial _{\tau }+\left[ \frac{m^{2}y^{2}}{4}+e^{m\tau }%
\right] \partial _{y}
\end{equation*}%
with conformal factors $\psi _{(1+2)}(\mathbf{Y}_{8})=\frac{2}{m}e^{\frac{m}{%
2}(\tau -x)}$, $\psi _{(1+2)}(\mathbf{Y}_{9})=-\frac{2}{m}e^{\frac{m}{2}%
(\tau +x)}$, and $\psi _{(1+2)}(\mathbf{Y}_{10})=\frac{m^{2}y}{2}$
respectively.

The conformally flat space%
\begin{equation}
d\bar{s}_{\left( 1+2\right) }^{2}=B^{2}\left( \tau \right) e^{2x}\left[
e^{m\tau}\left( -d\tau ^{2}+dx^{2}\right) +dy^{2}\right]  \label{b3.07}
\end{equation}%
admits the same elements of the conformal algebra with (\ref{b3.06}) but
with different conformal factors $\bar{\psi}_{\left( 1+2\right) }$. More
specifically it follows that 
\begin{equation}
\bar{\psi}_{(1+2)}(\mathbf{Y}_{A})=\mathbf{Y}_{A}\left[ \ln (Be^{x})\right]
+\psi _{(1+2)}(\mathbf{Y}_{A}).  \label{b3.08}
\end{equation}%
When we impose condition (\ref{sx2.1a}) we find that there does not exist
function $B\left( \tau \right) $ such that $\bar{\psi}_{(1+2)}(\mathbf{Y}%
_{A})$ to satisfy (\ref{sx2.1a}). On the other hand, we observe that for%
\begin{equation}
B\left( \tau \right) =e^{\mu \tau },\enskip\mu =\frac{m(\lambda -1)}{2}
\label{b3.09}
\end{equation}%
it follows$~\bar{\psi}_{(1+2)}(\mathbf{Y}_{7})=\lambda $, which means that $%
Y_{7}$ is reduced to a HV for \eqref{b3.07}. At this point it is important
to mention that $\bar{\psi}_{1+2}\left( \mathbf{Y}_{3}\right) =1$, however
there is only one proper HV and not two, as expected. We assume $\mathbf{Y}%
_{7}$ to be the proper HV and $\mathbf{Y}_{3}-\frac{1}{\lambda }\mathbf{Y}%
_{7}$ to be a KV.

For the four-dimensional decomposable spacetime%
\begin{equation}
ds_{(1+3)}^{2}=e^{2x}e^{2\mu \tau }\left[ e^{m\tau}\left( -d\tau
^{2}+dx^{2}\right) +dy^{2}\right] +dz^{2}  \label{b3.10}
\end{equation}%
from $Y_{7}$ we find the proper HV%
\begin{equation}
\mathbf{L}_{1}\equiv \mathbf{Y}_{7}+\lambda z\partial _{z}=\frac{2}{m}%
\partial _{\tau }+y\partial _{y}+\lambda z\partial _{z}.  \label{b3.11}
\end{equation}%
We conclude that the Bianchi III spacetime%
\begin{equation}
ds_{(III)}^{2}=e^{m\lambda \tau }A^{2}\left( \tau \right) \left( -d\tau
^{2}+dx^{2}+e^{-m\tau }dy^{2}+e^{-m\lambda \tau }e^{-2x}dz^{2}\right)
\label{b3.12}
\end{equation}%
admits the proper CKV $\mathbf{L}_{1}$ with conformal factor $\psi _{(III)}(%
\mathbf{L}_{1})=\frac{2}{m}\frac{A_{,\tau }}{A}+\lambda $ which reduces to a
HV when $A\left( \tau \right) $ is an exponential in which case the line
element is 
\begin{equation}
ds_{(III)}^{2}=-e^{m\kappa \tau }d\tau ^{2}+e^{m\kappa \tau
}dx^{2}+e^{m(\kappa -1)\tau }dy^{2}+e^{m(\kappa -\lambda )\tau }e^{-2x}dz^{2}
\label{b3.14}
\end{equation}%
or in equivalent form 
\begin{equation}
ds_{(III)}^{2}=-dt^{2}+\frac{m^{2}\kappa ^{2}t^{2}}{4}dx^{2}+\left( \frac{%
m^{2}\kappa ^{2}t^{2}}{4}\right) ^{\frac{\kappa -1}{\kappa }}dy^{2}+\left( 
\frac{m^{2}\kappa ^{2}t^{2}}{4}\right) ^{\frac{\kappa -\lambda }{\kappa }%
}e^{-2x}dz^{2}  \label{b3.15}
\end{equation}%
where now we write $\mathbf{L}_{1}=\kappa t\partial _{t}+y\partial
_{y}+\lambda z\partial _{z}$ with $\psi _{(III)}(\mathbf{L}_{1})=const\equiv
\kappa \neq 0$, recall that $dt=e^{\frac{m\kappa }{2}\tau }d\tau $.

Performing the same analysis for the second case of $\Gamma ^{2}\left( \tau
\right) =\kappa ^{-2}\cos ^{-2}\left( \tau \right) $ we find that the
resulting Bianchi III spacetime does not admit any proper CKV or a proper
HV, hence we omit the presentation of this analysis.

We summarize our results in the following proposition

\begin{proposition}
The only Bianchi III spacetime which admits a proper CKV is the 
\begin{equation}
ds^{2}=A^{2}\left( \tau \right) \left[ e^{m\lambda \tau } \left( -d\tau
^{2}+dx^{2}\right) + e^{m\left( \lambda -1\right) \tau} dy^{2} + e^{-2x}
dz^{2} \right].  \label{Bianchi3.1}
\end{equation}%
The CKV is $\mathbf{L}_{1}=\frac{2}{m}\partial _{\tau }+y\partial
_{y}+\lambda z\partial _{z}~$ and it has conformal factor $\psi _{(III)}(%
\mathbf{L}_{1})=\frac{2}{m}\frac{A_{,\tau }}{A}+\lambda $, where $A\left(
\tau \right) $ is an arbitrary function.
\end{proposition}

\section{Bianchi V spacetimes which admit a CKV}

\label{sec4}

For the computation of the CKVs for the Bianchi V spacetime we apply the
same procedure with Section \ref{sec3}, but for this case we start from the
two-dimensional spacetime%
\begin{equation}
ds_{\left( 2\right) }^{2}=\Gamma^2 \left( \tau \right) e^{-2x}\left(
-d\tau^{2}+dx^{2}\right) .  \label{b4.01}
\end{equation}%
The latter space is maximally symmetric only for $\Gamma^2 \left( \tau
\right) =e^{\gamma \tau }$ where the Ricci Scalar is calculated to be $%
R_{\left( 2\right) }=0$. It is important to mention that there does not
exist a function $\Gamma \left( \tau \right) $ such that the space (\ref%
{b4.01}) is of constant curvature.

We omit the intermediary calculations and we summarize the results in the
following proposition

\begin{proposition}
\label{pro2V} The Bianchi V spacetime 
\begin{equation}
ds^{2}=A^{2}\left( \tau \right) \left[ \Gamma ^{2}\left( \tau \right) \left(
-d\tau ^{2}+dx^{2}\right) +e^{2x}\left( B^{2}\left( \tau \right)
dy^{2}+dz^{2}\right) \right]  \label{Bianchi5}
\end{equation}%
admits the unique proper CKV $\mathbf{L}_{1}=\frac{2}{m}\partial _{\tau
}+y\partial _{y}+\lambda z\partial _{z}$ with $\psi _{(V)}(\mathbf{L}_{1})=%
\frac{2}{m}\frac{A_{,\tau }}{A}+\lambda $ only when $\Gamma ^{2}\left( \tau
\right) =e^{m\lambda \tau }, B^{2}\left( \tau \right) =e^{m(\lambda -1)\tau
} $. For $A^{2}\left( \tau \right) =e^{m(\kappa -\lambda )\tau }$ the CKV
reduces to a HV with conformal factor $\psi_{(V)}(\mathbf{L}%
_{1})=const=\kappa \neq 0$.
\end{proposition}

\section{Applications}

\label{sec5}

\subsection{Bianchi III cosmological fluid}

\label{sec5.1}

In this section we study some of the kinematic and the dynamic properties of
spacetime given by equation \eqref{b3.12} for the comoving observers $u^{a}=%
\frac{e^{-\frac{m\lambda }{2}\tau }}{A\left( \tau \right) }\delta _{\tau
}^{a}$ \textbf{,~} $u^{a}u_{a}=-1$. As it is well known (see e.g. \cite%
{ellis}) the four velocity of a class of observers introduces the 1+3
decomposition of tensor fields in spacetime. The decomposition of $u_{a;b}$
gives the kinematic quantities $\theta ,~\sigma ^{2},~\omega ^{2}$ and $%
\alpha ^{a}$ defined by the identity 
\begin{equation}
u_{a;b}=-\alpha _{a}u_{b}+\omega _{ab}+\sigma _{ab}+\frac{1}{3}\theta h_{ab}
\label{eq.1plus3u}
\end{equation}%
where $a^{a}=\dot{u}^{a}=u^{a}{}_{;b}u^{b}$, $\omega
_{ab}=h_{a}^{c}h_{b}^{d}u_{[c;d]}$, $\sigma _{ab}=\left( h_{a}^{c}h_{b}^{d}-%
\frac{1}{3}h^{cd}h_{ab}\right) u_{(c;d)}$, $\theta
=h^{ab}u_{a;b}=u^{a}{}_{;a}$, $\sigma ^{2}\equiv \frac{1}{2}\sigma
_{ab}\sigma ^{ab}$, $\omega ^{2}\equiv \frac{1}{2}\omega _{ab}\omega
^{ab},~h_{ab}=g_{ab}+u_{a}u_{b}.$ In this decomposition $a^{a}$ is the
4-acceleration of the observers $u^{a},$ and the quantities $\omega
_{ab},\sigma _{ab},\theta $ concerns the variation of the projected ($\bot
\delta x^{a}u_{a}=0$) connecting vector $\bot \delta x^{a}$ along the
congruence (i.e. the integral lines) defined by the vector field $u^{a}.$
The antisymmetric tensor $\omega _{ab}$ measures the relative rotation, the
tensor $\sigma _{ab}$ the anisotropic expansion and the scalar $\theta $ the
isotropic expansion of $\bot \delta x^{a}$. The dynamic variables of the
spacetime are defined by the 1+3 decomposition of the Einstein tensor $G_{ab}
$ as follows \textbf{{\cite{ellis}}} 
\begin{equation}
G_{ab}=\rho u_{a}u_{b}+2q_{(a}u_{b)}+ph_{ab}+\pi _{ab}  \label{eq.1plus3G}
\end{equation}%
\textbf{{where $\rho =G_{ab}u^{a}u^{b}$ is the energy-mass density of the
fluid, $p=\frac{1}{3}h^{ab}G_{ab}$ is the isotropic pressure, $%
q^{a}=-h^{ac}G_{cd}u^{d}$ is the heat flux tensor and $\pi _{ab}=\left(
h_{a}^{c}h_{b}^{d}-\frac{1}{3}h^{cd}h_{ab}\right) G_{cd}$ is the traceless
anisotropic tensor (measures the anisotropy of the fluid)}}.

Applying the above for the comoving observers in Bianchi III spacetime (\ref%
{b3.12}) we compute that the kinematic quantities $\omega ^{2}=0$, $%
\alpha^{a}=0$ while 
\begin{equation}
\theta =\frac{e^{-\frac{m\lambda }{2}\tau }}{A}\left[ 3\frac{d(\ln A)}{d\tau 
}+\frac{m(2\lambda -1)}{2}\right]  \label{b3.17}
\end{equation}%
and 
\begin{equation}
\sigma ^{2}=\frac{m^{2}(\lambda ^{2}-\lambda +1)}{12}\frac{e^{-m\lambda \tau
}}{A^{2}}.  \label{b3.18}
\end{equation}%
Similarly for the dynamic quantities we find that the (non-zero) components
for the cosmological fluid defined by the Bianchi III spacetime (\ref{b3.12}%
) are 
\begin{equation}
\rho =\frac{e^{-m\lambda \tau }}{4A^{2}}\left[ 4\frac{d\left( \ln A\right) }{%
d\tau }\left( 3\frac{d\left( \ln A\right) }{d\tau }+m\left( 2\lambda
-1\right) \right) +m^{2}\lambda \left( \lambda -1\right) -4\right]
\label{b3.19}
\end{equation}%
\begin{equation}
p=\frac{e^{-m\lambda \tau }}{A^{2}}\left[ -\frac{2}{A}\frac{d^{2}A}{d\tau
^{2}}+\frac{d\left( \ln A\right) }{d\tau }\left( \frac{d\left( \ln A\right) 
}{d\tau }+\frac{m}{3}(2-\lambda )\right) -\frac{m^{2}}{12}(\lambda
-1)(\lambda -2)+\frac{1}{3}\right]  \label{b3.20}
\end{equation}%
\begin{equation}
q^{a}=\left( 0,\frac{m\lambda }{2}\frac{e^{-\frac{3m\lambda }{2}\tau }}{A^{3}%
},0,0\right)  \label{b3.21}
\end{equation}%
\begin{equation}
\pi _{xx}=\frac{m(\lambda +1)}{3}\frac{d\left( \ln A\right) }{d\tau }+\frac{%
m^{2}(\lambda ^{2}-1)}{12}-\frac{1}{3}  \label{b3.22}
\end{equation}%
\begin{equation}
\pi _{yy}=e^{-m\tau }\left[ \frac{m(\lambda -2)}{3}\frac{d\left( \ln
A\right) }{d\tau }+\frac{m^{2}}{12}(\lambda -1)(\lambda -2)+\frac{2}{3}%
\right]  \label{b3.23}
\end{equation}%
and 
\begin{equation}
\pi _{zz}=-e^{-m\lambda \tau -2x}\left[ \frac{m(2\lambda -1)}{3}\frac{%
d\left( \ln A\right) }{d\tau }+\frac{m^{2}}{12}(\lambda -1)(2\lambda -1)+%
\frac{1}{3}\right] .  \label{b3.24}
\end{equation}

In the case of $A^{2}(\tau )=e^{m(\kappa -\lambda )\tau }$, where the CKV $%
\mathbf{L}_{1}$ becomes a HV, the above non-zero quantities are simplified
as follows 
\begin{equation}
\theta =\frac{m(3\kappa -\lambda -1)}{2}e^{-\frac{m\kappa }{2}\tau }
\label{b3.25}
\end{equation}%
\begin{equation}
\sigma ^{2}=\frac{m^{2}(\lambda ^{2}-\lambda +1)}{12}e^{-m\kappa \tau }
\label{b3.26}
\end{equation}%
\begin{equation}
\rho =\rho _{0}\left( m,\kappa ,\lambda \right) e^{-m\kappa \tau
}~,~p=p_{0}\left( m,\kappa ,\lambda \right) e^{-m\kappa \tau }  \label{b3.27}
\end{equation}%
\begin{equation}
q^{a}=\left( 0,\frac{m\lambda }{2}e^{-\frac{3m\kappa }{2}\tau },0,0\right)
\label{b3.28}
\end{equation}%
\begin{equation}
\pi _{xx}=\pi _{xx0}\left( m,\kappa ,\lambda \right) ~,~\pi _{yy}=\pi
_{yy0}\left( m,\kappa ,\lambda \right) e^{-m\tau }~,~\pi _{zz}=\pi
_{zz0}\left( m,\kappa ,\lambda \right) e^{-m\lambda \tau -2x}.  \label{b3.29}
\end{equation}

From the latter expressions we infer that for large $\tau$ and $m \kappa >0$
all the kinematical quantities, the mass density, the isotropic pressure and
the heat flux vector vanish. If, in addition, $\pi_{xx0}\left( m,\kappa
,\lambda \right)=0$, $m>0$ and $\lambda >0$, then for large $\tau $ the
fluid source vanishes and the solution describes an isotropic empty
spacetime.

\subsection{Bianchi V cosmological fluid}

\label{sec5.2}

We consider the extended Bianchi V spacetime of proposition \ref{pro2V} 
\begin{equation}
ds_{(V)}^{2}=-A^{2}(\tau )e^{m\lambda \tau }d\tau ^{2}+A^{2}(\tau
)e^{m\lambda \tau }dx^{2}+A^{2}(\tau )e^{m(\lambda -1)\tau
}e^{2x}dy^{2}+A^{2}(\tau )e^{2x}dz^{2}  \label{eq.biaV}
\end{equation}%
and repeat the calculations for the comoving observers. We find that the
kinematic quantities are exactly the same with those of the Bianchi III
spacetime while the dynamic (non-zero) dynamic variables of the cosmological
fluid are 
\begin{equation}
\rho =\frac{e^{-m\lambda \tau }}{4A^{2}}\left[ 4\frac{d\left( \ln A\right) }{%
d\tau }\left( 3\frac{d\left( \ln A\right) }{d\tau }+m\left( 2\lambda
-1\right) \right) +m^{2}\lambda \left( \lambda -1\right) -12\right]
\label{b4.19}
\end{equation}%
\begin{equation}
p=\frac{e^{-m\lambda \tau }}{A^{2}}\left[ -\frac{2}{A}\frac{d^{2}A}{d\tau
^{2}}+\frac{d\left( \ln A\right) }{d\tau }\left( \frac{d\left( \ln A\right) 
}{d\tau }+\frac{m}{3}(2-\lambda )\right) -\frac{m^{2}}{12}(\lambda
-1)(\lambda -2)+1\right]  \label{b4.20}
\end{equation}%
\begin{equation}
q^{a}=\left( 0,-\frac{m(\lambda +1)}{2}\frac{e^{-\frac{3m\lambda }{2}\tau }}{%
A^{3}},0,0\right)  \label{b4.21}
\end{equation}%
\begin{equation}
\pi _{xx}=\frac{m(\lambda +1)}{3}\frac{d\left( \ln A\right) }{d\tau }+\frac{%
m^{2}(\lambda ^{2}-1)}{12}  \label{b4.22}
\end{equation}%
\begin{equation}
\pi _{yy}=e^{-m\tau +2x}\left[ \frac{m(\lambda -2)}{3}\frac{d\left( \ln
A\right) }{d\tau }+\frac{m^{2}}{12}(\lambda -1)(\lambda -2)\right]
\label{b4.23}
\end{equation}%
and 
\begin{equation}
\pi _{zz}=-e^{-m\lambda \tau +2x}\left[ \frac{m(2\lambda -1)}{3}\frac{%
d\left( \ln A\right) }{d\tau }+\frac{m^{2}}{12}(\lambda -1)(2\lambda -1)%
\right].  \label{b4.24}
\end{equation}%
In the case $\mathbf{L}_{1}$ is a HV we deduce the same conclusions with the
Bianchi III case of section \ref{sec5.1}.

\section{Lie point symmetries of the wave equation}

\label{sec6}

Collineations of spacetimes can be used to construct symmetries and
conservation laws for some differential equations defined in curved
spacetimes. In \cite{mts1} it has been shown that there exists a unique
connection between the Noether symmetries for the geodesic Lagrangian of a
given Riemannian space and the elements of the admitted homothetic algebra.
Similar results have been shown for other partial differential equations of
special interest \cite{mts2,mts3}.

In this work we consider the wave equation 
\begin{equation}
\frac{1}{\sqrt{-g}}\frac{\partial }{\partial x^{\mu }}\left( \sqrt{-g}g^{\mu
\nu }\frac{\partial }{\partial x^{\nu }}\right) u\left( x^{\lambda }\right)
=0  \label{wave1}
\end{equation}%
in the Bianchi III spacetime (\ref{b3.12}) and in the Bianchi V spacetime (%
\ref{Bianchi5}) and determine its Lie symmetries. By following the generic
results of \cite{mts3}, we find that the admitted Lie point symmetries of
the wave equation in the Bianchi III spacetime (\ref{b3.12}) are the $3$
KVs, the vector field $Y_{u}=u\partial _{u}$, and the infinitely many
vectors $Y_{\infty }=b\left( x^{\mu }\right) \partial _{u}$ where $b\left(
x^{\mu }\right) $ is a solution of the original equation (\ref{wave1}). The
latter symmetry vector fields exist because equation (\ref{wave1}) is a
linear partial differential equation.

For a higher dimensional conformal algebra, equation (\ref{wave1}) admits
extra Lie point symmetries. Indeed, from our analysis and for the case where
the Bianchi III and Bianchi V spacetimes admit a proper HV the wave equation
becomes%
\begin{equation}
\left( -u_{tt}+u_{xx}+u_{yy}+e^{m\lambda t+2x}u_{zz}\right) +\frac{m}{2}%
\left( \lambda -2\kappa +1\right) u_{t}-u_{x}=0  \label{wave2}
\end{equation}%
or%
\begin{equation}
\left( -u_{tt}+u_{xx}+e^{mt-2x}u_{yy}+e^{m\lambda t-2x}u_{zz}\right) +\frac{m%
}{2}\left( \lambda +2\kappa -1\right) u_{t}+4u_{x}=0.  \label{wave3}
\end{equation}

Then we find that equation (\ref{wave2}) admits the generic Lie point
symmetry vector%
\begin{equation}
Y_{III}=\left( a_{1}\frac{2}{m}\right) \partial _{t}+a_{2}\partial
_{x}+\left( a_{1}y+a_{3}\right) \partial _{y}+\left( a_{1}\lambda
z+a_{2}z+a_{4}\right) \partial _{z}+\left( a_{u}u+a_{\infty }b\left(
t,x,y,z\right) \right) \partial _{u}  \label{wave4}
\end{equation}%
while equation (\ref{wave3}) is invariant under the one parameter point
transformation with generator%
\begin{equation}
Y_{V}=\left( a_{1}\frac{2}{m}\right) \partial _{t}+a_{2}\partial _{x}+\left(
a_{1}y-a_{1}y+a_{3}\right) \partial _{y}+\left( a_{1}\lambda
z-a_{2}z+a_{4}\right) \partial _{z}+\left( a_{u}u+a_{\infty }b\left(
t,x,y,z\right) \right) \partial _{u}.  \label{wave5}
\end{equation}

The latter symmetry vectors can be applied to construct conservation laws or
similarity solutions for the wave equation. However, such an analysis is
beyond the scope of the present work.

\section{Conclusion}

\label{sec7}

In this paper we have shown that there is only one type of Bianchi III and
Bianchi V spacetime given respectively in (\ref{b3.12}) and (\ref{Bianchi5})
which admit a single proper CKV. Furthermore, two more spacetimes are found
which admit a HV. In order to arrive at this result we applied an algorithm
which relates the CKVs of decomposable spacetimes with the collineations of
the non-decomposable subspace. The kinematics of the fluid of the comoving
observers in all these four spacetimes is not accelerating and rotating and
has only expansion and shear a result compatible with the anisotropy of the
Bianchi spacetimes. Concerning the dynamics it has been shown that the fluid
of these observers is heat conducting and anisotropic, that is it is a
general fluid. Finally we have used the conformal vectors we found in each
case in order to determine the generators of the Lie symmetries of the wave
equation in the Bianchi III spacetime (\ref{b3.12}) and in the Bianchi V
spacetime (\ref{Bianchi5}).

\subsection*{\textbf{Acknowledgement}}

AP thanks the University of Athens for the hospitality provided while this
work was carried out.

\end{document}